\documentclass{emulateapj}

\newcommand{\grb}{GRB~150424A}
\newcommand{\pks}{PKS~J0949$-$2511}
\newcommand{\swift}{\textit{Swift}}
\newcommand{\dm}{\ensuremath{{\rm pc}\,{\rm cm}^{-3}}}
\newcommand{\pasa}{PASA}

\slugcomment{\apj\ Letters, in press}

\shorttitle{Limits on Prompt Radio Emission from GRB~150424A}
\shortauthors{Kaplan et al.}

\def\CASS{{2}}
\def\Curtin{{9}}
\def\USydney{{8}}
\def\UWisc{{1}}
\def\CAASTRO{{3}}
\def\berkeley{{6}}
\def\eureka{{7}}
\def\uva{{4}}
\def\astron{{5}}
\begin{document}

\title{A Deep Search for Prompt Radio Emission from the Short GRB
  150424A With The Murchison Widefield Array}

\author{D.~L.~Kaplan\altaffilmark{\UWisc},
  A.~Rowlinson\altaffilmark{\CASS,\CAASTRO,\uva,\astron},
  K.~W.~Bannister\altaffilmark{\CASS,\CAASTRO},
  M.~E.~Bell\altaffilmark{\CASS,\CAASTRO},
  S.~D.~Croft\altaffilmark{\berkeley,\eureka},
  T.~Murphy\altaffilmark{\USydney,\CAASTRO},
  S.~J.~Tingay\altaffilmark{\Curtin,\CAASTRO},
  R.~B.~Wayth\altaffilmark{\Curtin,\CAASTRO}, 
A.~Williams\altaffilmark{\Curtin},
}

\altaffiltext{\UWisc}{Department of Physics, University of
  Wisconsin--Milwaukee, Milwaukee, WI 53201, USA; \email{kaplan@uwm.edu}}

\altaffiltext{\CASS}{CSIRO Astronomy and Space Science (CASS), PO Box
  76, Epping, NSW 1710, Australia}
\altaffiltext{\CAASTRO}{ARC Centre of Excellence for All-sky
  Astrophysics (CAASTRO)}
\altaffiltext{\uva}{Anton Pannekoek Institute for Astronomy,
  University of Amsterdam, Science Park 904, 1098 XH Amsterdam, The
  Netherlands}
\altaffiltext{\astron}{ASTRON, The Netherlands Institute for Radio Astronomy, Postbus 2, 7990 AA, Dwingeloo, The Netherlands}
\altaffiltext{\berkeley}{Astronomy Department, University of California, Berkeley,
  501 Campbell Hall \#3411, Berkeley, CA 94720, USA}
\altaffiltext{\eureka}{Eureka Scientific, Inc., 2452 Delmer Street
  Suite 100, Oakland, CA 94602, USA}
\altaffiltext{\USydney}{Sydney Institute for Astronomy, School of Physics, The University of Sydney, NSW 2006, Australia}
\altaffiltext{\Curtin}{International Centre for Radio Astronomy
  Research, Curtin University, Bentley, WA 6102, Australia}

\begin{abstract}
We present a search for prompt radio emission associated with the
short-duration gamma-ray burst (GRB) 150424A using the Murchison
Widefield Array (MWA) at frequencies from 80--133\,MHz.  Our
observations span delays of 23\,s--30\,min after the GRB,
corresponding to dispersion measures of 100--7700\,\dm.  We see no
excess flux in images with timescales of 4\,s, 2\,min, or 30\,min, and
set a  3\,$\sigma$ flux density limit of 3.0\,Jy at 132\,MHz on the
shortest timescales: some of the most stringent limits to
date on prompt radio emission from any type of GRB.  We use these
limits to constrain a number of proposed models for coherent emission
from short-duration GRBs, although we show that our limits are not
particularly constraining for fast radio bursts because of reduced
sensitivity for this pointing.  Finally, we discuss the prospects for
using the MWA to search for prompt radio emission from gravitational
wave transients and find that while the flux density and luminosity
limits are likely to be very constraining, the latency of the
gravitational wave alert may limit the robustness of any conclusions.
\end{abstract}


\section{Introduction}
The Advanced LIGO interferometers
\citep[aLIGO;][]{2015CQGra..32g4001T} have very recently started observational science
runs, soon to be joined by other upgraded detectors.  For the first
time there is a realistic prospect for detection of an astrophysical
gravitational wave (GW) transient, with a range of possible
electromagnetic counterparts \citep{2012ApJ...746...48M}.  Rapid
multi-wavelength followup might then allow detection and
characterization of astrophysical gravitational wave sources (see
e.g.,
  \citealt{2014ApJ...795..105S,2014ApJ...789L...5K}),
greatly enhancing the scientific utility of such a discovery.
For instance, we might be able to conclusively determine the origin of
short-duration GRBs (SGRBs; see
\citealt{2014ARAA..52...43B} and \citealt{2015arXiv150902922F} for recent reviews)
which are generally accepted to
originate from neutron star-neutron star mergers.

Even before aLIGO begins operation, prompt radio followup of SGRBs may 
give clues as to their origin and help tie them to other mysterious
phenomena.  Specifically, a number of authors have suggested the
possibility of prompt, coherent radio emission right before, during or
right after neutron star-neutron star mergers through a variety of
physical mechanisms \citep[e.g.,][]{2000AA...364..655U,2010ApSS.330...13P}.
This may serve as an explanation \citep{2013PASJ...65L..12T,2014ApJ...780L..21Z,2014AA...562A.137F} for Fast Radio Bursts
\citep[FRBs;][]{2007Sci...318..777L,2013Sci...341...53T}:
impulsive ms bursts of dispersed radio emission with peak flux
densities of $\sim 1$\,Jy or more at 1.4\,GHz and apparently cosmological origins.

Searches for prompt radio emission from GRBs have been conducted for
decades
but most have concentrated on the more-common long-duration GRBs
(LGRBs) and/or not been very sensitive (see
\citealt{2014ApJ...785...27O} and \citealt{2014ApJ...790...63P} for
recent discussions). 
Observations that covered the times of the GRBs were usually from
less-sensitive all-sky instruments
\citep[e.g.,][]{1996MNRAS.281..977D,2014ApJ...785...27O}, while more
sensitive pointed observations often took several minutes to slew before
starting to observe
\citep[e.g.,][]{2012ApJ...757...38B,2014ApJ...790...63P}.  
We instead take advantage of the capabilities of the Murchison
Widefield Array (MWA;
\citealt{2009IEEEP..97.1497L,2013PASA...30....7T}) --- a low frequency
(80--300\,MHz) interferometer located in Western Australia --- for
rapid, sensitive followup. With fully-electronic steering and a wide
field-of-view, it can respond to astrophysical transients within 20\,s
of receiving an alert as we demonstrate below.

\begin{figure*}
\plotone{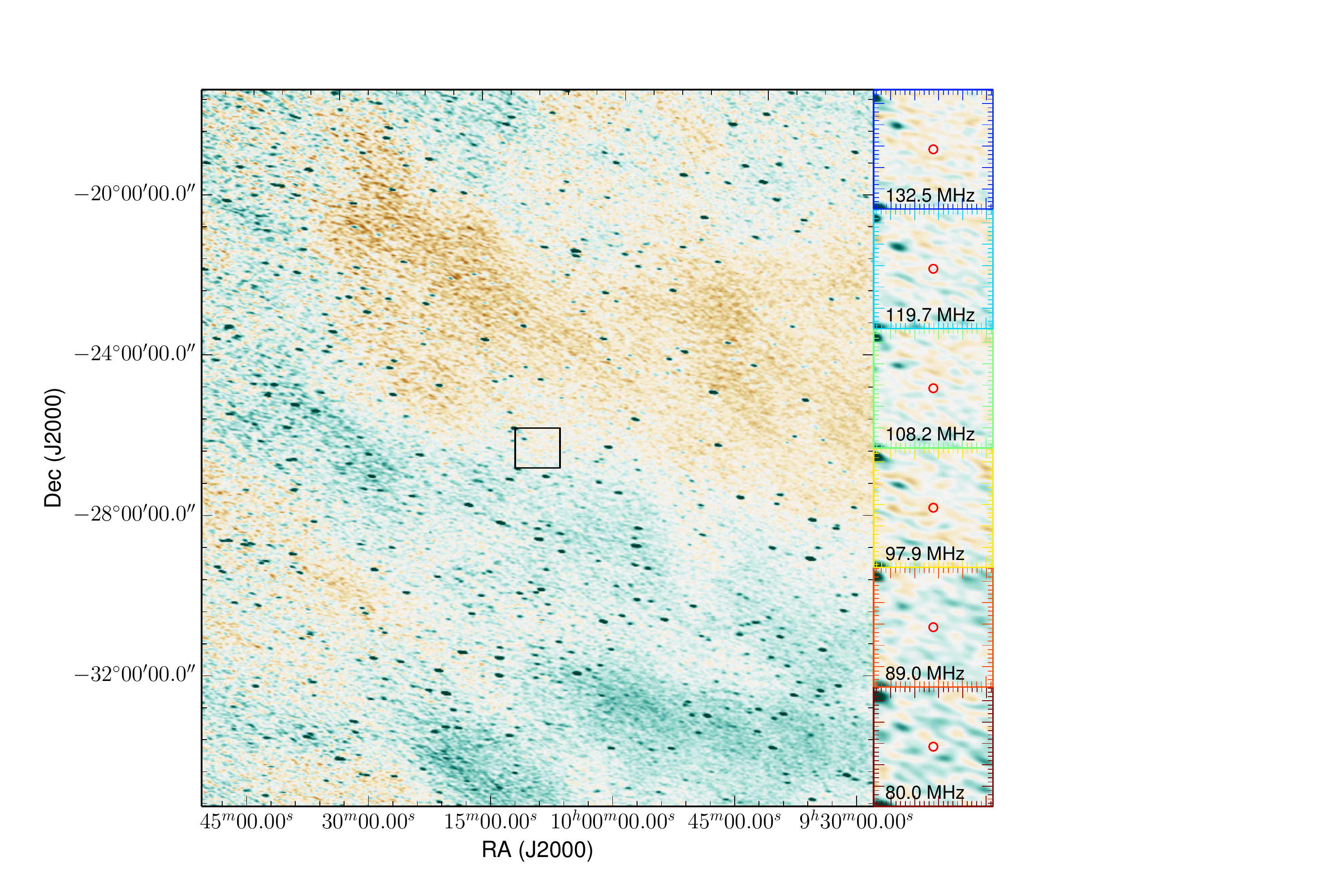}
\caption{MWA image of the field of \grb.  We show a $15\degr \times
  15\degr$ portion of the 30\,min mosaic in the 132.5\,MHz sub-band.
  The box shows the sizes of the $1\degr \times 1\degr$ insets on the
  right, each of which shows the same portion of the field but for
  each sub-band (as labeled).  The position of the GRB is indicated by
  the circle, and is known to $\ll 1\,$pixel.  }
\label{fig:image}
\end{figure*}

Here we present a search for prompt low-frequency radio emission
associated with the short-duration \grb\ using the MWA.  \grb\ was detected on
2015 April 25 at 07:42:57\,UT by the Burst Alert Telescope (BAT) on
board the \swift\ satellite
\citep{2004ApJ...611.1005G,2015GCN..17743...1B}. The $\gamma$-ray
emission consists of  multiple very bright pulses with a duration of about
0.5\,s, followed by weak $\gamma$-ray emission up to 100\,s after the
initial pulses \citep{2015GCN..17761...1B}. 
\grb\ is thus classified as a SGRB with
extended emission (EE SGRB): a small population of GRBs whose
properties are most consistent with SGRBs despite their long durations
(e.g., \citealt{2010ApJ...717..411N,2011ApJ...735...23N}), and where
the origin of the extended emission is still being debated but may
involve a magnetar central engine
\citep[e.g.,][]{2008MNRAS.385.1455M,2014MNRAS.438..240G}.  The X-ray
Telescope (XRT) began observing the location of \grb\ 87.9\,s after
the burst and found a bright, fading X-ray source.  Followup
observations \citep{2015GCN..17758...1C} identified a redshift $z =
0.3$ galaxy $5\arcsec$ (projected separation of 22.5\,kpc) away from
the optical afterglow \citep{2015GCN..17745...1P}.  However,
\citet{tanvir+15} found a fainter potential host galaxy with a likely
redshift of $z >0.7$ underlying the GRB location.  We note that the
density of the medium surrounding this GRB is unknown and, if high,
may impede the detection of coherent radio emission
\citep{2007ApJ...658L...1M}.

All cosmological quantities in this paper are computed based on
\citet{2014AA...571A..16P}.  We use a nominal redshift of 0.7 for our
calculations, consistent with \citet{tanvir+15}.

\section{Observations \& Analysis}
The MWA Monitor and Control computer received a socket-based notice
from the Gamma-ray Coordinate Network (GCN)
at 07:43:10\,UT and  quickly
scheduled 30\,min of observations of \grb.
To save time, the telescope stayed in
the same configuration as the previous observation which had been
solar observing.  This used an unusual configuration with the 24
coarse 1.28\,MHz channels spread out in a ``picket fence,'' mode, with
2.56\,MHz sub-bands spread between 80\,MHz and 240\,MHz (and using
0.5\,s correlator integrations with 40\,kHz frequency resolution).
Observations started at 07:43:20\,UT, 23\,s after the GRB.  This was
during the day at the MWA (Sun at $25\degr$ elevation) and with the
GRB somewhat low in the sky (elevation $30\degr$), although it was
$123\degr$ away from the Sun.  Because of the low elevation the MWA
had less sensitivity and a more irregular primary  beam shape than usual.  The
observations consisted of 15 individual 112\,s scans, separated by 8\,s.

\begin{figure*}
\plotone{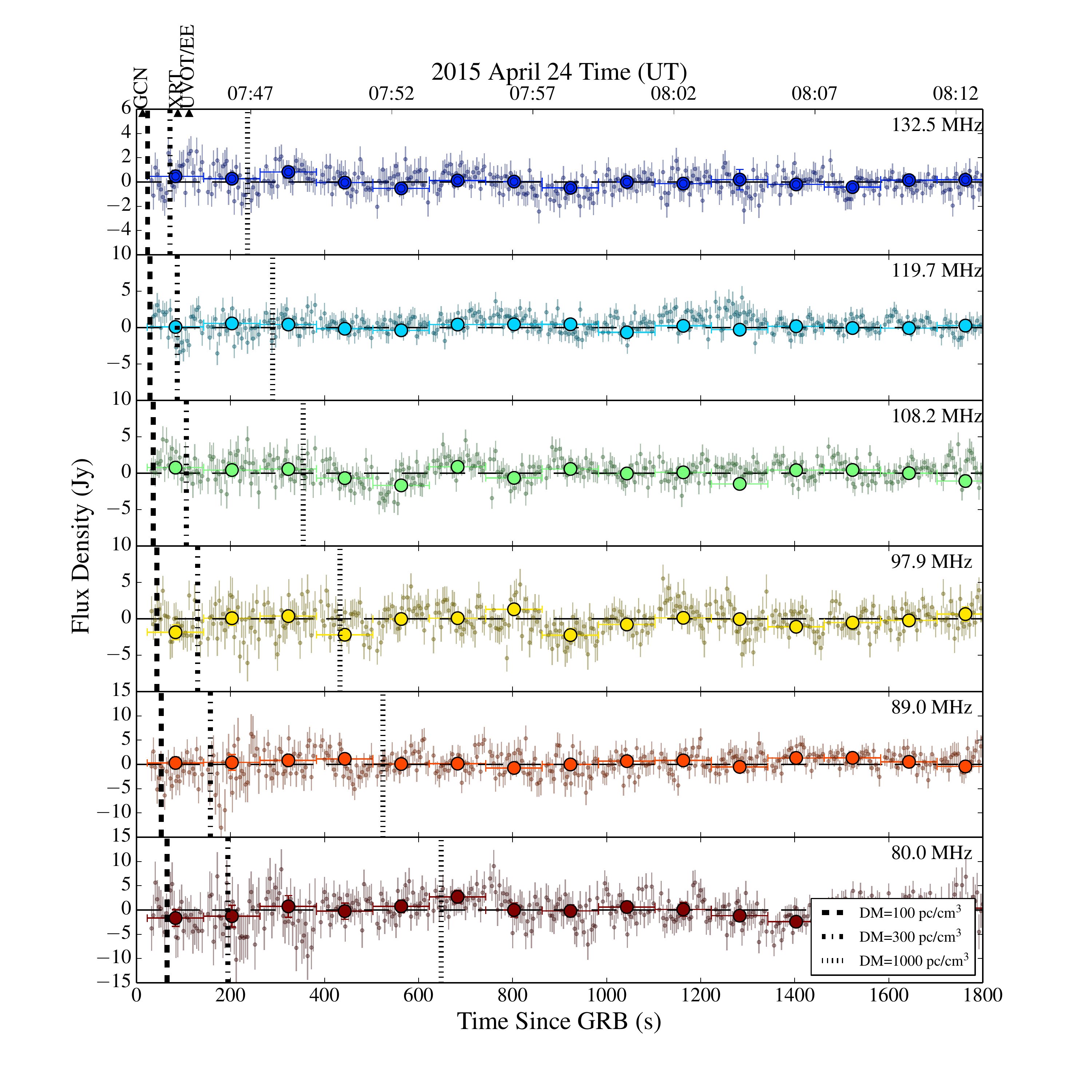}
\caption{Flux densities at the position of \grb\ in each sub-band.  We
  show measurements from the individual 4\,s images (points) as well
  as 2\,min images (circles).  The black triangles in the top left
  corner show the times of the GCN, XRT, and UVOT observations as
  labeled; the time of the UVOT observation was also roughly the end
  of the extended emission (EE) period. We also show the appropriate
  delays relative to the time of the GRB for DMs of 100\,\dm\ (dashed
  vertical lines), 300\,\dm\ (dot-dashed vertical lines) and
  1000\,\dm\ (dotted vertical lines).}
\label{fig:fluxes}
\end{figure*}

The processing followed standard MWA procedures
\citep[e.g.,][]{2014PASA...31...45H}.  
We performed initial phase calibration using an observation of Hydra~A
taken earlier in the same day in the same mode. We then imaged the
scans with $4096\times 4096$ $0\farcm6$ pixels in the XX and YY
instrumental polarizations using \texttt{WSClean}
\citep{2014MNRAS.444..606O}, using 40,000 \texttt{CLEAN} iterations
and allowing for one round of amplitude and phase self-calibration (as
demonstrated by \citealt{rowlinson15},
this does not remove transients
as long as they do not dominate the total flux density of the image).
Finally, we corrected the instrumental polarization to Stokes I (total
intensity) using the primary beam from \citet{2015RaSc...50...52S}.
The synthesized beam was elongated with an axis-ratio of 2.6:1 because
of the low elevation; the major axis varied from $12\arcmin$ to
$4\farcm2$ over the different sub-bands.  Examining the images from
the different sub-bands, the upper 6 sub-bands (frequencies $\geq
144\,$MHz) suffered significant image artifacts, mostly due
to uncleaned sidelobes from Hydra~A ($18\degr$ to the north west of
 \grb) and primary beam grating lobes that encompassed the Sun.  We
ended up discarding the upper 6 sub-bands as we could not
satisfactorily improve the image quality.  For the remaining
sub-bands, we combined individual 2\,min scans into a single 30\,min
mosaic (as in \citealt{2014PASA...31...45H}); we show the mosaics for
each sub-band in Figure~\ref{fig:image}.  The flux density scale was
corrected so that the bright, unresolved source \pks\ ($4\fdg5$ away
from GRB) averaged over each 2\,min observation matched the spectral
energy distribution we interpolated from values from the NASA Extragalactic Database, given in Table~\ref{tab:limits}.
We
then also created images with 4\,s integration times, using the
corrected $uv$ data but only performing 100 \texttt{CLEAN} iterations
on each.

\begin{figure}
\plotone{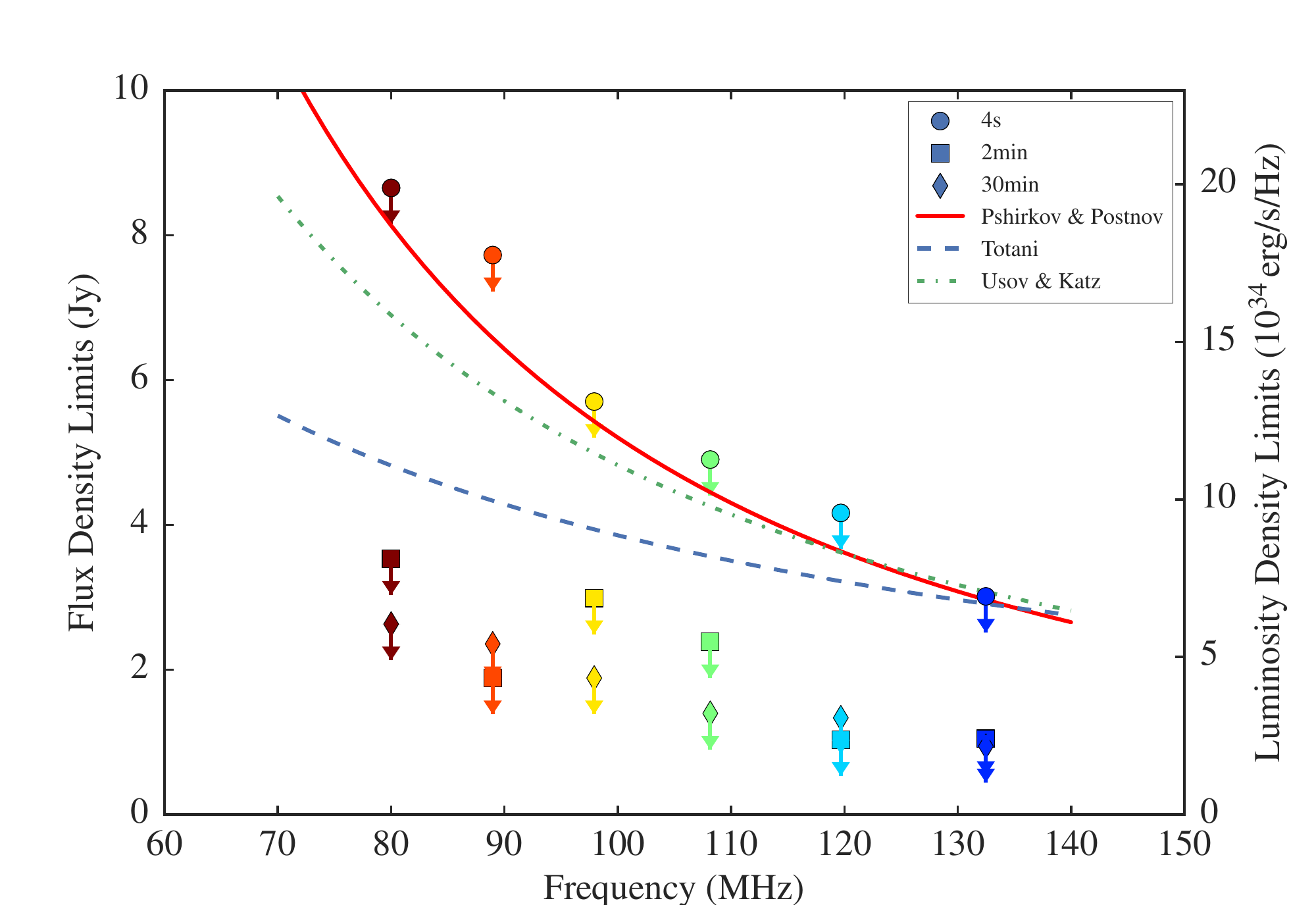}
\caption{Flux density limits (3\,$\sigma$) for \grb\ in each sub-band,
  based on the 4\,s images (circles), 2\,min images (squares), and
  30\,min mosaics (diamonds); also see Table~\ref{tab:limits}.  We also show the luminosity density
  limits appropriate for a redshift $z=0.7$; if the redshift is
  instead 0.3 (1), then the luminosity densities limits would decrease
  (increase) by a factor of 7 (2.4). We also show example predictions
  that have been adjusted to not exceed our 4\,s limits from
  \citet[][with $\dot E \lesssim 5 \times 10^{50}\,{\rm erg\,s}^{-1}$
    and assuming an efficiency scaling exponent
    $\gamma=0$]{2010ApSS.330...13P}, \citet[][with efficiency
    $\epsilon_r\lesssim 5\times 10^{-2}$, magnetic field
    $B=10^{13}\,$G and initial spin period
    $P=0.5\,$ms]{2013PASJ...65L..12T}, and \citet[][with efficiency
    $\delta \lesssim 3.5\times 10^{-7}$]{2000AA...364..655U} appropriate
  for ${\rm DM}<444\,$\dm.}
\label{fig:limits}
\end{figure}

\begin{deluxetable*}{l c c c c c c}
\tablewidth{0pt}
\tablecaption{Reference Flux Densities and 3-$\sigma$ Limits\label{tab:limits}}
\tablehead{
 & \colhead{80.0\,MHz} & \colhead{88.9\,MHz} & \colhead{97.9\,MHz} & \colhead{108.1\,MHz} & \colhead{119.7\,MHz} & \colhead{132.5\,MHz}}
\startdata
Flux Density of \pks\ (Jy)\dotfill & 22.5 & 21.8 & 21.1 & 20.1 & 19.1 & 17.9 \\
4\,s Flux Density Limits for \grb\ (Jy)\dotfill  & \phn8.7 & \phn7.7 & \phn5.7 & \phn4.9 & \phn4.2 & \phn3.0 \\
2\,min Flux Density Limits for \grb\ (Jy)\dotfill  & \phn3.5 & \phn1.9 & \phn3.0 & \phn2.4 & \phn1.0 & \phn1.1 \\
30\,min Flux Density Limits for \grb\ (Jy)\dotfill  & \phn2.6 & \phn2.4 & \phn1.9 & \phn1.4 & \phn1.3 & \phn0.9 
\enddata
\end{deluxetable*}

For each set of images: 4\,s, 2\,min, and 30\,min mosaics, we measured
the flux density of \pks\ along with the flux density at the position
of the GRB (position uncertainty $\ll 1$\,pixel; we verified that the
position variation of \pks\ due primarily to ionospheric refraction
was $\lesssim 1\,$pixel) and the image noise properties.  In
Figure~\ref{fig:fluxes} we show the flux densities at the position of
\grb\ for each sub-band from both the 4\,s and 2\,min images.  There
is some degree of correlation between individual points
\citep{2014MNRAS.438..352B}, but as a whole the data are noise-like
with reduced $\chi^2$ values near 1 (0.76--0.98 depending on the
band).  We searched for statistically significant peaks in each of the
sub-bands over a range of timescales from 4\,s--2\,min and see nothing
exceeding 3\,$\sigma$, much less anything that is correlated between
the sub-bands (with a possible delay allowing for interstellar
dispersion).  We then determined 3\,$\sigma$ flux density limits, shown in
Figure~\ref{fig:limits} and Table~\ref{tab:limits}.  Note that the
88.9 and 119.7\,MHz sub-bands are slightly anomalous in that the
limits from the 30\,min mosaics are slightly worse than those from
2\,min images.  This may be from a combination of source confusion
limiting the sensitivity of the mosaics and residual poorly-cleaned
sidelobes from Hydra~A.  As a whole, though, the 4\,s sub-bands behave
well, and the limits from the longer integrations are lower, almost by
the factor of 5 expected from the integration time.

\section{Discussion}
In our discussion of \grb, we consider how our observations constrain
the potentially related phenomena of SGRBs and FRBs, and furthermore
the implications of these results on low-frequency radio follow-up of
GW transients.
But first, we need to address the effects on any radio signal of
propagation through intervening ionized media.

\subsection{Propagation Effects}
Any prompt radio signal from \grb\ is expected to be modified by its
propagation through the interstellar medium (ISM) of its host galaxy,
the intergalactic medium (IGM), and the ISM of the Milky Way
\citep{2007ApJ...658L...1M}.  Free electrons will introduce
dispersion, causing lower frequencies to arrive later
while inhomogeneities will cause scattering that smears out temporal
structure.  Dispersion is quantified by the dispersion measure (DM):
the integral of the line-of-sight electron density.  We can expect a
DM of about 80\,\dm\ from the Milky Way \citep{2002astro.ph..7156C},
and perhaps a roughly similar contribution from the GRB's host galaxy.
We expect the DM from the IGM to be roughly $1000\,z\,\dm$ for a
redshift $z$ \citep{2004MNRAS.348..999I,2013ApJ...776L..16T}, so we
can expect DM$_{\rm IGM}=300\,$\dm--1000\,\dm\ depending on the actual
redshift of the GRB, and a total DM of 500\,\dm--1200\,\dm.  In
Figure~\ref{fig:fluxes} we plot the time delays in each sub-band for a
range of DMs.  Even for the lowest possible DMs (just the Milky Way)
our observing covered the delayed time of any prompt signal,
especially in the lower sub-bands.  Our 30\,min observation spans the
nominal DM range quite well, and we sample up to a DM of
2800\,\dm\ for the lowest sub-band or 7700\,\dm\ for the highest.
Note that the dispersion across a bandpass of 2.56\,MHz would last
9\,s--40\,s depending on the sub-band for a nominal DM of 1000\,\dm,
so a fast pulse would last 2--10 of our 4\,s images.  We assume that
scattering does not significantly smear out any signal
\citep{2013Sci...341...53T,2013MNRAS.436L...5L,2013ApJ...776..125M},
but note that this may need to be revisited as more information is
gained about FRB behavior.

\subsection{Short-Duration Gamma-Ray Bursts}
Given the observed SGRB, we can constrain any associated prompt,
coherent radio signal such as those predicted in models of neutron
star-neutron star mergers
\citep[e.g.,][]{2010ApSS.330...13P,2013PASJ...65L..12T} or more
generic GRB phenomenon \citep[e.g.,][]{2000AA...364..655U}.  These
models have poorly-predicted efficiency factors which we are able to
constrain directly from our observations. We show example predictions
that have been adjusted to not exceed our 4\,s limits in
Figure~\ref{fig:limits}.  {For the rapid magnetized spin-down model of
\citet{2010ApSS.330...13P}, we have spin-down luminosity $\dot E
\lesssim 5 \times 10^{50}\,{\rm erg\,s}^{-1}$ and assume an efficiency
scaling exponent $\gamma=0$, while for the similar but lower $B$ model of
\citet{2013PASJ...65L..12T}, we have efficiency $\epsilon_r\lesssim
5\times 10^{-2}$, along with nominal magnetic field $B=10^{13}\,$G and
initial spin period $P=0.5\,$ms. And for coherent radio emission from
the magnetized wind of a magnetar central engine colliding with the
ambient medium as in \citet{2000AA...364..655U}, we have ratio of
radio to $\gamma$-ray fluence $\delta \lesssim 3.5\times 10^{-7}$.
Note that our constraints here are for a fixed observed timescale of
4\,s, which limits the DM to 444\,\dm\ for 133\,MHz observations.  At
higher DMs, our constraints will scale up accordingly.  These
constraints will be explored further in  Rowlinson et al.\ (2015b in prep).}
{With
  a detection we can use the fluence, duration,
  and delay of any coherent emission to strongly constrain any model.}

\begin{figure}
\plotone{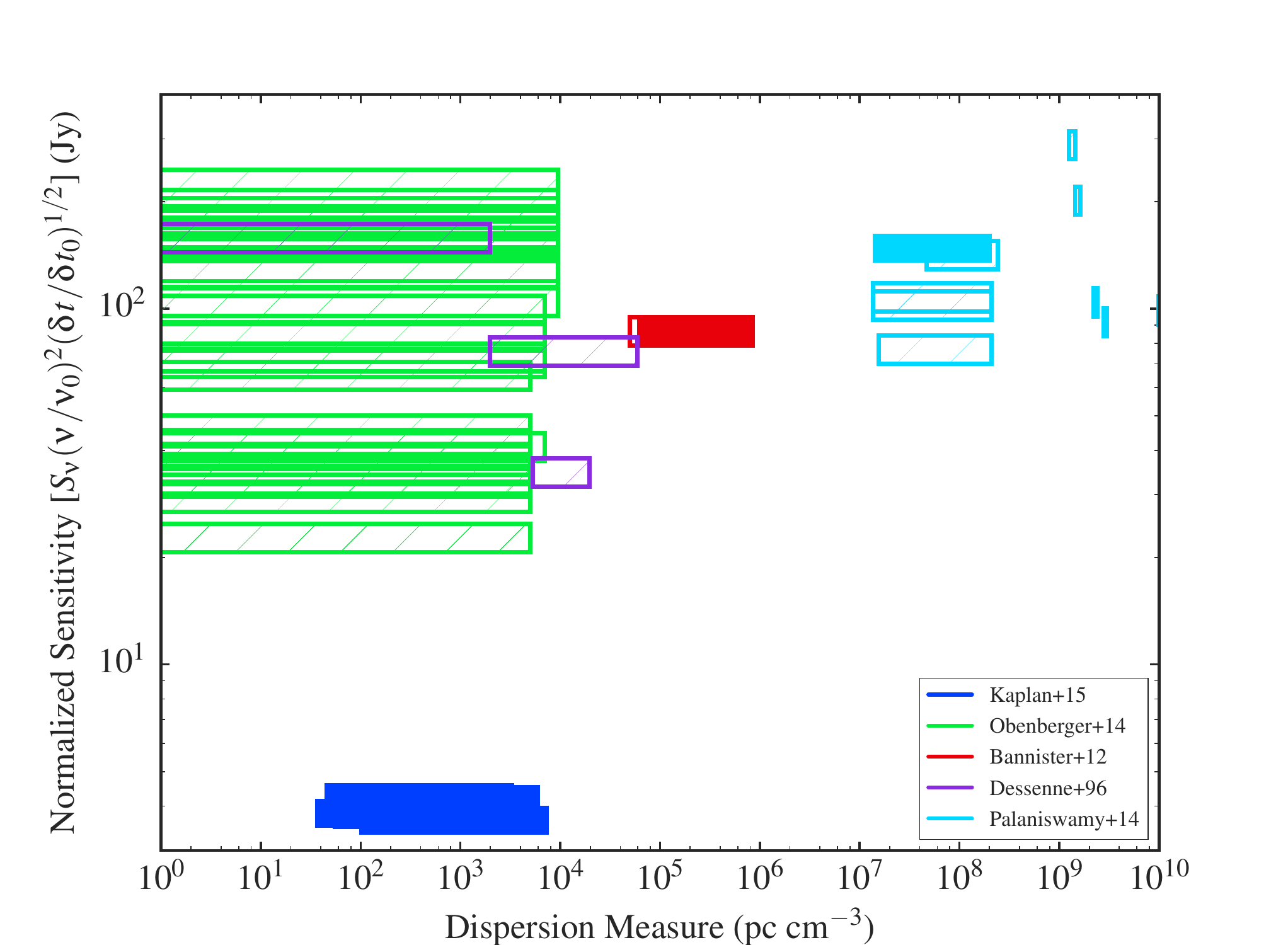}
\caption{Limits to prompt emission from GRBs, based on
  \citet[][38--74\,MHz; green]{2014ApJ...785...27O},
  \citet[][1400\,MHz; red]{2012ApJ...757...38B}, \citet[][151\,MHz;
    purple]{1996MNRAS.281..977D}, \citet[][2300\,MHz;
    cyan]{2014ApJ...790...63P}, and this paper (blue).  The follow-up
  times have been converted into an effective dispersion measure,
  assuming that the radio emission is coincident with the GRB.  The
  flux density limits have been normalized by frequency (assuming
  $S_\nu \propto \nu^{-2}$; e.g., \citealt{2010ApSS.330...13P}) to
  $\nu_0=100$\,MHz and to a common timescale of $\delta t_0=10$\,s.
  Short-duration GRBs --- those listed as such in the literature ---
  are filled shapes, while long-duration GRBs are hatched.  The same
  GRB can be shown multiple times if it was observed at different
  telescopes/frequencies.  {For comparison, FRBs are detected at DMs of
  400--1100\,\dm\  with  1.4-GHz peak flux densities of $\sim 1\,$Jy
  over a $\sim 10\,$ms pulse \citep{2015MNRAS.447.2852K}}: {scaled to a
  10-s observation this would be a factor of $\sim 10^3$ too low to
  display here.}}
\label{fig:comparison}
\end{figure}

In Figure~\ref{fig:comparison} we compare our observations to other GRB
searches from the literature.  To compare observations at a range of
frequencies and timescales, we convert them to a common sensitivity
assuming $S_\nu \propto \nu^{-2}$ (e.g., \citealt{2010ApSS.330...13P})
and that sensitivity scales as $1/\sqrt{\delta t}$ (with $\delta t$
the integration time).  We see that our limits are a factor of $\sim
10$--100 deeper than those from \citet[][assuming no
  detections]{2012ApJ...757...38B} or \citet{2014ApJ...785...27O}, and
cover far closer to the time of the GRB than the former.

\subsection{Fast Radio Bursts}
Some of the models for FRBs tie them directly to neutron star-neutron
star mergers and SGRBs \citep[e.g.,][]{2013PASJ...65L..12T,2014ApJ...780L..21Z}.
For example, \citet{2014ApJ...780L..21Z} predict a FRB when a magnetar
central engine powering the GRB collapses to form a black hole, which
might happen at the end of the extended emission phase \citep[][but
  see \citealt{2014MNRAS.438..240G}]{2015ApJ...805...89L}. Since our
observations cover from right after the GRB (allowing for dispersion)
to well past the end of the extended emission, we can place the first
constraints on this model for the extended emission.

In our most sensitive sub-bands of 133\,MHz, we set a 3\,$\sigma$
limit to the flux density of any short-duration emission of
$<3.0$\,Jy.  This translates into a fluence limit of $<12.0\,{\rm
  Jy}\,$s, compared to FRB fluences at 1.4\,GHz of $<1$\,Jy\,ms to
$>30$\,Jy\,ms \citep{2015MNRAS.447.2852K}.  Assuming flux densities
scale $S_\nu \propto \nu^{\alpha}$, we can only exclude FRBs with
spectral indices $\alpha < -2.5$.  This is not particularly constraining
\citep[unlike][]{2015MNRAS.452.1254K,tingay15,rowlinson15}, largely because of
the reduced sensitivity of the MWA at this low elevation
\citep[cf.][]{2013ApJ...776L..16T} and with the contribution of the
Sun to the system temperature.  It is also possible that the
1.4-GHz FRB detections have been aided by interstellar scintillation
\citep{2015MNRAS.451.3278M} which would not help at these
frequencies.

\subsection{Gravitational Wave Transients}
Finally, we can consider the constraints on GW transients.  The aLIGO
detectors were not operating during \grb, so no direct GW limit can be
determined, but we can consider the prospects for MWA followup of GW
transients.
As discussed in
\citet{2014ApJ...795..105S}, the error regions for GW triggers in
2015--2016 can cover hundreds of square degrees. 
Moreover, they need not be compact or simply connected.  While the
nominal field-of-view of the MWA is about $600\,{\rm deg}^2$ at
150\,MHz, we cannot always cover all of the expected error regions.
Unless the GW event occurs within the MWA's field-of-view (chance of
$\approx 1$\%), we will need to re-point following a GW trigger.

Given the expected range of redshift/DM for GW events (intergalactic
DMs of 10--50\,\dm, or total DMs of 50--200\,\dm), we expect time
delays from the GW event of only $41 (\nu/100\,{\rm MHz})^{-2} ({\rm
  DM}/100\,\dm)\,$s, not including possible internal delays
\citep{2014ApJ...780L..21Z}.  As we have demonstrated, 20\,s is
sufficient for MWA followup, but the bigger question is the latency of
the GW detection and notice.  Currently the low-latency compact binary
coalescence pipeline is expected to send out notices with a time delay
of 90--120\,s after the GW event
\citep{2014ApJ...795..105S,T1400054-v7}, although this could decrease
as the signal-to-noise increases, with a detection potentially even
occurring before the merger \citep{2012ApJ...748..136C}. Although this
can be mitigated at some level by moving to frequencies $\lesssim
60\,$MHz where the dispersive delay increases to surpass the GW event
delay this delay may ultimately be a significant
limitation for the prospects of prompt GW followup \citep{2015arXiv150906876C}.

If we are able to point appropriately, we expect a limiting flux
density of about 0.1\,Jy, or luminosity limits of $10^{38-39}\,{\rm
  erg\,s}^{-1}$ for typical distances.  Since GW sources would be at
redshifts $<0.05$ compared to 0.3--1 here, any radio emission would be
significantly brighter by a factor of 50--1000.  {This would lead to
much more realistically constraining models for FRBs and SGRBs, with
e.g., $\epsilon_r$ from \citet{2013PASJ...65L..12T} close to the value
of $10^{-4}$ seen for radio pulsars, or an $\dot E$ from
\citet{2010ApSS.330...13P} close to the range inferred from modeling
extended emission in SGRBs \citep{2015MNRAS.448..629G}.}

\section{Conclusions}
We have demonstrated prompt followup with a pointed radio telescope
that we have used to set stringent limits to any prompt, coherent
emission from the short \grb.  Looking on our fastest timescale of
4\,s, we set 3\,$\sigma$ flux density limits of 3.0\,Jy at 133\,MHz.
These limits are a factor of $\sim 100$ lower that most prior limits,
and cover delays of 23\,s--30\,min after the GRB, corresponding to DMs
of 100--7700\,\dm.  We did not detect any FRB coincident with the GRB,
but these limits are not very constraining compared to the population
of FRBs because of reduced sensitivity for this particular pointing.

We plan to continue our GRB followup program over the next year,
although given the preferred elevation range of $>45\degr$ the rate of
\swift\ SGRBs suitable for MWA followup is $<1\,{\rm yr}^{-1}$.
However, this serves as a demonstration and template analysis for
future followup of GW transients --- particularly timely given the
very recent start of science runs with the aLIGO detectors.  We will
work to improve the analysis time for the MWA data to facilitate
multi-wavelength followup over the large GW error regions
\citep{2014ApJ...795..105S}.  Additional work in reducing the latency
of the GW triggers will also be helpful since that is expected to be a
limitation on the robustness of any conclusions from low-frequency
radio searches.

\acknowledgements We thank an anonymous referee for useful comments
and J.-P.~Macquart, C.~Trott, A.~Urban, A.~Offringa, and S.~B.~Cenko
for helpful discussions.  This  work uses  the
Murchison Radio-astronomy Observatory, operated by CSIRO. We
acknowledge the Wajarri Yamatji people as the traditional owners of
the Observatory site.  Support for the operation of the MWA is
provided by the Australian Government Department of Industry and
Science and Department of Education (National Collaborative Research
Infrastructure Strategy: NCRIS), under a contract to Curtin University
administered by Astronomy Australia Limited. We acknowledge the iVEC
Petabyte Data Store and the Initiative in Innovative Computing and the
CUDA Center for Excellence sponsored by NVIDIA at Harvard University.
DLK and SDC are additionally supported by NSF grant AST-1412421.  This
research made use of APLpy, an open-source plotting package for Python
hosted at \url{http://aplpy.github.com}.  This research has made use
of the NASA/IPAC Extragalactic Database (NED), which is operated by
the Jet Propulsion Laboratory, California Institute of Technology,
under contract with the National Aeronautics and Space Administration.

{\it Facilities:} \facility{Murchison Widefield Array}.

\bibliographystyle{apj} 

\end{document}